\documentclass[11pt,twoside]{article}
\usepackage{./asp2010}

\resetcounters

\def\kms{\rm\,km\ s^{-1}}
\def\vsini{$v\sin i$}
\def\mmax{$m_{\rm max}$}

\begin{document}

\title{Completing the Massive Star Population:  Striking Into the Field}
\author{M. S. Oey and J. B. Lamb}
% PLEASE DO NOT CHANGE THE APPEARANCE OF MY NAME WITHOUT 
%     CONSULTING ME, THANKS.
\affil{University of Michigan\\
Astronomy Department, 830 Dennison, Ann Arbor, MI, 48109-1042, USA}

\begin{abstract}
As a population, field massive stars are relatively enigmatic, and
this review attempts to illuminate this sector of the 
high-mass stellar population, which comprises 20 -- 25\% of the massive
stars in star-forming galaxies.  The statistical properties of the
field population are vital diagnostics of star formation
theory, cluster dynamical evolution, and stellar evolution.
We present evidence that field massive
stars originate both {\it in situ} and as runaways from clusters,
based on the clustering law, IMF, rotation velocities, and individual
observed {\it in situ} candidate field stars.  We compare the known
properties of field and cluster massive stars from studies in the
Magellanic Clouds and the Galaxy, including our RIOTS4 complete
spectroscopic survey of SMC OB stars.  In addition to the origin of
the field massive stars, we discuss additional properties including
binarity, runaway mechanisms, and some evolved spectral types.
\end{abstract}

\section{Introduction:  Definition and frequency of field OB stars}

In the last session, we discussed massive stars within clusters,
and I now turn to a more enigmatic component of the high-mass
stellar population:  the field stars.  What is the nature of the field
massive star population?  In particular, are there OB stars that form
{\it in situ} in the field, more or less isolated?  Or are most
massive field stars runaways that are ejected from clusters?
Note that ``field'' stars are subject to different
definitions, especially regarding the degree of isolation, both in
the number of immediate companion stars and distance from massive
clusters.  Thus what follows is a review of studies that adopt
necessarily heterogeneous definitions of the field.

For a simple definition of field OB stars, we consider the clustering
law, which is the number of OB stars $N_*$ per cluster.  This is
simply a manifestation of the cluster mass function, and both are
known to robustly follow a --2 power-law distribution, which for the
clustering law is:
\begin{equation}\label{eq_clustering}
N(N_*)\ dN_* \propto N_*^{-2}\ dN_* \quad .
\end{equation}
If we consider only the OB stars and ignore all the lower-mass stars,
then we find that this power-law extends smoothly down to $N_*=1$,
i.e., individual OB stars, which we define to be field stars \citep{Oey2004}.  
We obtained this result using the $UBVR$ photometric survey of
the entire Small Magellanic Cloud (SMC) by \citet{Massey2002},
uniformly identifying all OB star candidates from the criteria that
the reddening-free parameter $Q_{UBR} \leq -0.84$ and $B\leq 15.21$.
Following \citet{Battinelli1991}, we applied a friends-of-friends algorithm
to assign stars to clusters, adopting the clustering length
that maximizes the number of identified clusters, which is 28 pc.
OB candidates having no other OB candidate within this distance are
then defined to be field stars.

That these field stars are simply a smooth extension of the power-law
clustering suggests that they are dominated by objects that are simply
the ``tip of the iceberg'' on small, sparse clusters containing only
the single OB stars that formed {\it in situ} as members of these groups.
However, note that since the clustering law is presented as a
logarithmic relation, we can also accommodate a significant
contribution, even 50\%, from runaway stars, and still present a
smooth power law for the clustering law.

From this power law (equation~\ref{eq_clustering}), the fraction of
field massive stars is (Oey et al. 2004):
\begin{equation}\label{eq_fieldfrac}
f({\rm field}) = \big(\ln N_{*,\rm up} + 0.5772\big)^{-1} \quad ,
\end{equation}
where $N_{*,\rm up}$ is the number of OB stars in the richest cluster
in the entire galaxy or ensemble.  For most astrophysical situations,
$f({\rm field})$ is in the range of about 1/3 to 1/10.  For the SMC,
  equation~\ref{eq_fieldfrac} predicts $f(\rm field)\sim 0.20$, while
the observed value is about 0.26.  This is comparable to the estimated
frequency of $\sim$20\% for O stars in the Milky Way \citep{Mason2009,
Gies1987}.

\section{The Field IMF}

Given this uniformly identified sample of field OB candidates for the
entire SMC, we are in the final stages of a survey to obtain
spectroscopic observations for all of these stars:  the Runaways and
Isolated O-Type Star Spectroscopic Survey of the SMC (RIOTS4).  This
is being carried out at the Magellan Baade 6.5-m telescope by
multislit observations with the IMACS imaging spectrograph at a
resolution of $R\sim 2600$.  This yields spectroscopic classifications
and radial velocities good to 10 -- 15 $\kms$.  Our coverage of the SMC
bar region is essentially complete, while the SMC wing region is
currently about 85\% complete.

A primary goal of our survey is thus to provide a definitive
measurement of the upper IMF in the field, which is presented
at this meeting by Joel Lamb.  As is well known, Massey (2002) and
\citet{Massey1995} found an upper IMF power-law slope that is much
steeper than the Salpeter value $\Gamma = - 1.35$ in both the Large
and Small Magellanic Cloud from observations of limited areas in these
galaxies, obtaining $\Gamma\sim -4$.  Our preliminary results confirm
a steeper slope, though with a less extreme value: $\Gamma\sim -3$
(Lamb et al. 2011, these proceedings).  Similarly, \citet{vandenBergh2004}
evaluated the distribution of Milky Way O stars as a function
of spectral type and found a preponderance toward late types in the
field relative to associations.  This is thus consistent with a
steeper field slope in our Galaxy as well.  

Since Wolf-Rayet (WR) stars have higher-mass progenitors than typical
OB stars, we can also examine the relative frequency of WR stars to
probe the field and cluster IMF.  In the SMC, two WR stars are in our
field sample, out of a total of 12 known WR stars in this galaxy
\citep{MasseyDuffy2001, Massey2001}.
This yields a field WR frequency of 17\% relative to 26\% for OB
stars.  While this difference is not statistically significant because
of the tiny number of WR stars, we note that the results do suggest
a lower WR frequency, as expected for a steeper field IMF. 

\section{Oe, Be, and B[e] Stars}

Classical Oe and Be stars are another type of massive star
that are easy to identify from their strong Balmer emission lines.
However, the origin and frequency of this phenomenon is less
straightforward than for WR stars, since apparently only some fraction of
OB stars go through an Oe/Be phase.  Field vs cluster properties for
these stars are best examined in the Magellanic Clouds, but no clear
differences have been identified in the distributions of spectral
type, luminosity class, and rotation velocities \citep{Keller1999,
Martayan2006, Martayan2007}.  Evidence for variation in their
frequency between field and clusters is also inconclusive.  Our RIOTS4
survey yields a preliminary frequency of 53\% and 21\% for field Be
and Oe stars, respectively.  These high frequencies reflect the known
prevalence of Oe/Be stars at low metallicity for the SMC
\citep[e.g.,][]{Maeder1999, Martayan2006}. 

We identified a new B[e] supergiant in the course of the RIOTS4
survey, the star LHA115-S62 \citep{Henize1956}.  Thus we have
one B[e] supergiant in the field relative to the total of 6 in the SMC,
adding to the five previously known B[e] supergiants \citep{Wisniewski2007}.
This yields a frequency of 17\% B[e] supergiants in the field, which
is the same value found for WR stars.

\section{Binary Stars}

Statistics for binary stars are an especially important diagnostic for
the origin of field stars, since theories for both star formation and
dynamical ejection from clusters can make specific predictions for the
frequencies and binary parameters for field stars.  Because of the
difficult and time-consuming nature of binary identification across
the vast parameter space in binary mass ratio, separation,
inclination, and period, different techniques are needed to probe the
full parameter space \citep[e.g.,][]{SanaEvans2011, Mason2009},
ranging from high-resolution imaging with speckle interferometry to
high-resolution spectroscopic monitoring.  Furthermore, obtaining reliable
statistics also depends on observing complete samples over appropriate
time domains.  Mason et al. (2009) have
compiled the most complete statistics for massive star binaries to
date, examining  347 Milky Way O stars in a magnitude-limited sample
catalogued by \citet{Maiz2004}.  To date, they
find that the field and cluster O stars have frequencies of 59\% and
75\%, respectively, that are at least binary or with greater
multiplicity.  While this difference is statistically significant,
it is important to bear in mind that binary frequencies for both
populations are lower limits.  Sana \& Evans (2011) find a cluster
binary frequency around 50\% for 9 Galactic OB associations; since
these are typically farther than the stars in the Ma\' iz-Apell\'aniz
et al. sample, the binary detection rate is lower.  The binary
frequencies for their extragalactic clusters are lower still.
The frequency of binary runaway O stars is 43\% \citep{Mason2009}.
%, based on *** pairs.

Bragan\c ca et al. (2011, in preparation) obtained observations of 379
solar neighborhood B stars with the MIKE echelle spectrograph at
Magellan.  Preliminary results show that about 21\% of these stars are 
double-lined spectroscopic binaries, with analysis of field vs cluster
members in progress.  Our RIOTS4 survey of the SMC includes binary
monitoring of three fields, corresponding to about 10\% of the sample.  With
observations of a dozen epochs in hand, about 9 out of 16 non-Be
stars are single-lined spectroscopic binaries, corresponding to 60\%
of the sample.  We caution that stochasticity is an important factor
in this preliminary result.

As mentioned above, the statistics for binary stars promise to be
critical diagnostics for the origin of both {\it in situ} and runaway
field stars.  At present, reliable statistics for binary orbital
parameters have not yet been published, but we stress the importance of
obtaining these data.  For example, we almost have useful
estimates of the non-compact binary runaway frequency, which is a
critical probe of dynamical ejection (\S 6).   

\section{Rotation Velocities}

Rotation velocities are the parameter for which the field and cluster
stars show the most revealing differentiation.  It has long been known
that field stars show lower values of \vsini\ than cluster stars 
\citep[e.g.,][]{Guthrie1984}.  This difference is a critical clue
to the origin of field stars, and the proposed models reflect the
different possible mechanisms for these origins.  \citet{Wolff2007,
Wolff2008} suggest that the different rotational velocities result from
the different star-formation environments, with lower turbulence in
small star-forming regions resulting in lower \vsini\ relative to
high-density, intense star-forming regions.  However, Huang and
collaborators suggest that the lower \vsini\ in the field is simply
linked to a higher average age for field stars, which may be dispersed
from clusters, reflecting an evolutionary spin-down effect
\citep{HuangGies2008, Huang2010}.  They 
demonstrate that both field and cluster stars show an 
anticorrelation between \vsini\ and $\log g$, confirming this
spin-down.  

\begin{figure}[!ht]
\plotone{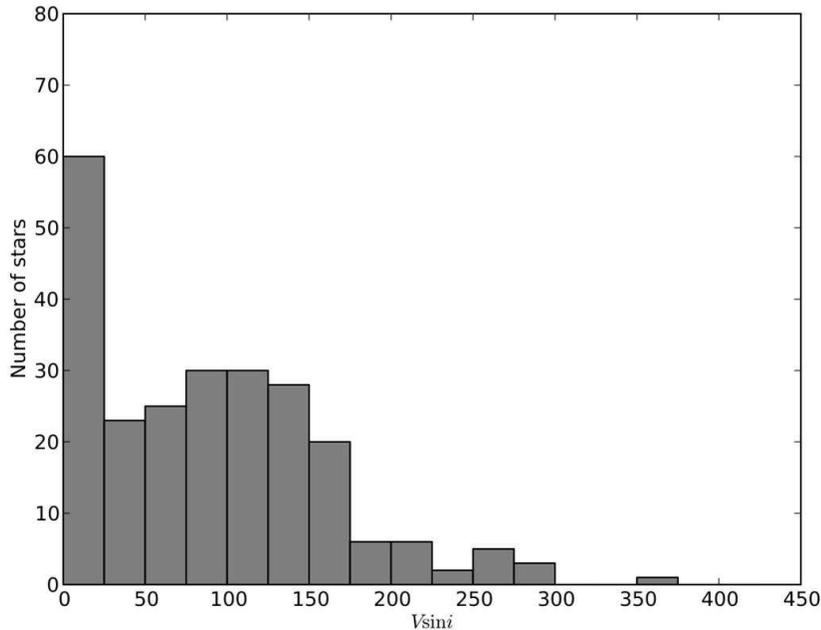}
\caption{
Preliminary distribution of \vsini\ for solar neighborhood B stars
(Bragan\c ca et al., in preparation).
\label{fig_braganca}}
\end{figure}

However, the field \vsini\ distribution strongly appears to be
bimodal.  We observed a large sample of solar neighborhood B stars
with the MIKE echelle spectrograph at Magellan, and removed
known cluster members, yielding 240 remaining stars (Bragan\c ca et
al. 2011, in preparation).  Interestingly, this only heightens the
bimodality apparent in the \vsini\ distribution
(Figure~\ref{fig_braganca}).  The bimodality is also evident in the field 
star sample of \citet{Abt2002}, selected by \citet{Strom2005} and
plotted by Wolff et al. (2007).  There are also hundreds of stars in
this sample, so the double-peak form of the \vsini\ distribution is
statistically significant, and similar to that in Figure~\ref{fig_braganca}.
Garmany et al. (2010) also identified a bimodal distribution in
their sample of 150 Milky Way outer disk B stars, which were also
observed with MIKE at Magellan.  In all cases, we
see a component of slow rotators which peak below \vsini\ $\sim25\
\kms$, and fast rotators with a much broader \vsini\ component peaking
near 100 $\kms$.  Meanwhile, a sample of high-latitude, field
main-sequence B stars studied by \citet{Martin2006} does {\it not} show the
bimodal distribution and apparently shows only the broad component of
fast rotators.  Martin argues that since these stars are outside
the Galactic plane, they must be strongly dominated by runaway stars.
Therefore, we suggest that {\it the fast rotators seen in the bimodal
distributions are runaway stars, while the population of slow rotators
are field stars that formed in situ.}  This is consistent with the
model of Wolff et al. (2007) that \vsini\ correlates with star
formation density.  Therefore, this may be strong evidence
that massive stars do form {\it in situ} in the field.

\section{Runaway Stars}

Runaway stars ejected from clusters are commonly observed and
understood to be a significant component of the field massive star
population.  O stars show a higher frequency, around 10 -- 30\%
runaways, relative to B stars, around 5 -- 10\% runaways \citep{Moffat1998,
Gies1987}.  This mass dependence is an important
clue to the origin of runaway stars, for which there are two general
models.  The first is dynamical ejection from 3- and 4-body
interactions, with the latter taking place as binary-binary
interactions, in which the ejected stars may or may not be separated.
There has been substantial work done in simulating these dynamical
interactions, pioneered by, e.g., \citet{Poveda1967}, \citet{Mikkola1983},
and \citet{LeonardDuncan1988, LeonardDuncan1990}.  The second mechanism to generate
runaway stars is via supernova ejection, in which a binary companion
explodes and releases the surviving star as a runaway \citep[e.g.,]
[]{Zwicky1957, Blaauw1961}.

It is difficult to determine which of these mechanisms dominates, but
that result would itself be a vital diagnostic for cluster dynamical
evolution, binary fractions, and star formation.  The strongest test
is the frequency of runaway, double-lined spectroscopic 
binaries (SB2) or visual binaries, which {\it must} be generated by the
dynamical ejection scenario.  Note that single-lined spectroscopic
binaries may have a compact secondary, which could be generated by the
supernova ejection model.  Out of 42 runaways in the 
Galactic field O star sample, Mason et al. (2009) identify 3 as SB2
and 11 showing visual multiplicity.  The two categories are
not mutually exclusive, and the visual multiples may include
line-of-sight interlopers.  Thus, the frequency of non-compact, multiple
runaways may be around 10 -- 20\% of O star runaways.  If this relatively
high value is confirmed, it demonstrates the significance of the dynamical
ejection mechanism.  This may even dominate the massive runaway population,
since this process also contributes many single runaway stars.

\citet{Blaauw1993} suggested that stars ejected by the supernova binary
mechanism should be He-enriched from contamination by the explosion
event.  Such an enhancement indeed has been observed by Blaauw and by
\citet{Hoogerwerf2001} 
in a sample of local runaway stars at distances within 700
pc.  Another diagnostic, proposed by Martin (2006), is that the
runaway star's \vsini\ should correlate with its space velocity,
since the star's runaway velocity should be closely linked to its
former binary orbital velocity.  To produce a high, runaway velocity,
the orbital velocity also must have been high, implying a tight binary
that was most likely orbitally locked with the companion, and hence
the ejection and orbital velocities should be similar.  However, Martin's
(2006) sample of high-latitude field B stars does not show such a
correlation, suggesting that the supernova ejection mechanism may not
be important for that population.  While Martin (2006) comes to the opposite
conclusion from Hoogerwerf et al. (2001), he
notes that the two studies are based on very different samples, with
the latter perhaps representing a younger runaway population, owing to
the stars' proximity within the Galactic disk.  Further studies can examine
the age-dependence of runaway frequencies and properties by
targeting evolved stars \citep[e.g.,][]{BergerGies2001, Slettebak1997}.
Our understanding is still extremely uncertain, but the modest
existing evidence described above perhaps slightly favors dynamical ejection
over supernova ejection as the dominant mechanism for producing
massive runaway stars.

\begin{figure}[!ht]
%\vspace*{-1cm}
\plotfiddle{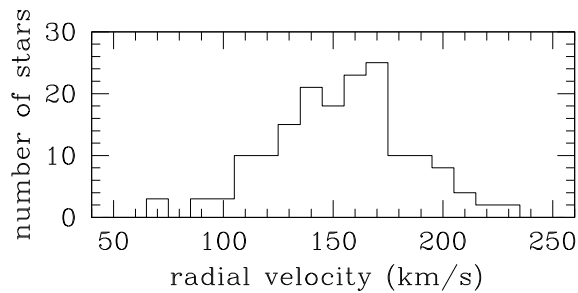}{3cm}{0}{40}{40}{-120}{-10}
\caption{
Preliminary distribution of RIOTS4 radial velocities for SMC field OB stars.
\label{fig_RV}}
\end{figure}

Much more study is needed, especially observations to better quantify the
properties of runaway stars.  With our RIOTS4 survey, we are able to
measure radial velocities of our complete OB star sample with typical
accuracy of 10 -- 15 $\kms$.  Our preliminary distribution
(Figure~\ref{fig_RV}) has a FWHM
$\sim 60\ \kms$, which includes large-scale SMC motions.  In particular,
the SMC Wing region is offset from the SMC Bar by about 30 $\kms$.
From the fairly smooth velocity distribution, it is therefore
difficult to infer the existence of any significant features 
that may originate from different runaway mechanisms, but quantitative
predictions are needed to fully exploit this complete data set.  Our
results are qualitatively similar to the distributions presented by
\citet{EvansHowarth2008}, who compiled the radial velocities for samples of
OB stars that include both cluster and field stars.  Our sample is
smaller than theirs, but with much stronger selection criteria and 
spectral resolution that is almost double theirs.

\section{In-Situ Field Stars}

The existence of massive field stars that formed {\it in situ} may
depend on the exact definition of field stars (\S 1).  While
candidates for such stars
do exist (e.g., Selier et al., these proceedings), it remains to
be determined whether any of these form in true isolation.  But in
any case, the sparsest regime sets strong
constraints on the physics of star formation.  For example,
it is much more difficult to generate isolated massive stars with
competitive accretion models \citep[e.g.,][]{Bonnell2004} than with
core collapse models \citep[e.g.,][]{KrumholzMcKee2008}.

There do appear to be examples of massive stars in sparse groups
containing only a few other low-mass stars.  Such groups might be
remnants of rapidly-dissolved clusters, but it remains likely that
these are examples of extreme, low-density star formation.  \citet{deWit2004,
deWit2005} find that locally, about 12\% of Galactic O stars are
in such sparse groups.  Using {\sl HST}/ACS observations, we found three
SMC field O stars to be within sparse, low-mass groups, out of six
apparently-isolated field O stars \citep{Lamb2010}.  We carried out
Monte Carlo simulations based on the --2 power law clustering
(equation~\ref{eq_clustering}) and a \citet{Kroupa2001} IMF, assuming
these two relations to be completely independent; in particular, we
assumed no dependence of the stellar upper-mass limit on cluster
mass.  The observed sparse O star groups fall easily within the allowed
parameter space of these simulations when examining the mass ratio of
the two highest-mass stars and the maximum-mass star \mmax\ per cluster.
However, these sparse O star groups are inconsistent with the proposed
relation between \mmax\ and cluster mass, suggesting that in general, the
integrated galaxy IMF (IGIMF) should not be steeper than the IMF in
individual clusters.

\section{Conclusion:  In-Situ Field Stars In-Situ and Runaways}

The evidence presented above supports the existence of both
runaways and an {\it in-situ} component contributing significantly to
the field massive star population.  We commonly observe runaway stars
and we know that they exist; thus, the main question is whether field
massive stars also form {\it in situ}.  I suggest that several lines
of evidence support their existence.  First, there are many instances
of observed high-mass stars that are in sparse, low-mass groups, or
even candidate {\it in situ} isolated stars.  Second, the bimodal
\vsini\ distribution strongly suggests that the slow rotators are {\it
  in situ} field OB stars.  Third, the upper IMF in the field appears
to be significantly steeper than measured in clusters; whereas, if the field
consists purely of runaways, then its upper IMF should be flatter than
in clusters, since O stars have a higher runaway frequency than B stars.
And fourth, there is no discontinuity between the field stars and
clusters in the cluster mass function (equation~\ref{eq_clustering}).

The runaways originate from either dynamical ejection or supernova
ejection, and presently, the evidence does not strongly favor one
scenario over the other.  There may be slightly stronger support for
dynamical ejection, since we do see a non-negligible frequency of
double-lined spectroscopic binaries and visual binaries, both of which
can only originate by dynamical ejection.  The lack of correlation
thus far, between rotation and space velocity also suggests that 
supernova ejection is not a dominant process.  However, observed
He abundance enhancements in runaway stars do support the supernova
ejection model.  All these results are based on limited,
individual studies and far more work is needed to establish the trends,
and better probe diagnostics of the ejection mechanisms.

Our understanding of the field massive star population is gradually
emerging, but much more study is needed to fully illuminate this
enigmatic, but substantial, population.  An especially critical area
is the binary population:  some of the most powerful diagnostics will
be revealed by statistics of the binary orbital parameters.  These are
difficult and time-consuming to obtain, but they will be powerful
diagnostics of runaway ejection mechanisms, constrain star formation
and IMF theories, and cluster evolution, as well as the nature of the
field population.  Binary population synthesis models and N-body
dynamical simulations are already mature areas of study, and more
observations in particular are needed to test the models.

\acknowledgements 

I gratefully acknowledge the organizers for conference logistical 
support; and thanks to Andrew Graus for RIOTS4 statistics and for
Figure~\ref{fig_RV}.  This work was supported by grants NSF
AST-0907758 and NASA HST-GO-10629.01. 

\bibliography{Oey_Tonyfest}

\begin{thebibliography}{}
\expandafter\ifx\csname natexlab\endcsname\relax\def\natexlab#1{#1}\fi
\expandafter\ifx\csname url\endcsname\relax
  \def\url#1{\texttt{#1}}\fi
\expandafter\ifx\csname urlprefix\endcsname\relax\def\urlprefix{URL }\fi
\providecommand{\eprint}[2][]{\url{#2}}

\bibitem[{{Abt} et~al.(2002){Abt}, {Levato}, \& {Grosso}}]{Abt2002}
{Abt}, H.~A., {Levato}, H., \& {Grosso}, M. 2002, \apj, 573, 359

\bibitem[{{Battinelli}(1991)}]{Battinelli1991}
{Battinelli}, P. 1991, \aap, 244, 69

\bibitem[{{Berger} \& {Gies}(2001)}]{BergerGies2001}
{Berger}, D.~H., \& {Gies}, D.~R. 2001, \apj, 555, 364

\bibitem[{{Blaauw}(1961)}]{Blaauw1961}
{Blaauw}, A. 1961, Bull.Astron.Inst.Netherlands, 15, 265

\bibitem[{{Blaauw}(1993)}]{Blaauw1993}
--- 1993, in Massive Stars: Their Lives in the Interstellar Medium, edited by
  {J.~P.~Cassinelli \& E.~B.~Churchwell}, vol.~35 of ASP Conf. Series, 207

\bibitem[{{Bonnell} et~al.(2004){Bonnell}, {Vine}, \& {Bate}}]{Bonnell2004}
{Bonnell}, I.~A., {Vine}, S.~G., \& {Bate}, M.~R. 2004, \mnras, 349, 735

\bibitem[{{de Wit} et~al.(2004){de Wit}, {Testi}, {Palla}, {Vanzi}, \&
  {Zinnecker}}]{deWit2004}
{de Wit}, W.~J., {Testi}, L., {Palla}, F., {Vanzi}, L., \& {Zinnecker}, H.
  2004, \aap, 425, 937

\bibitem[{{de Wit} et~al.(2005){de Wit}, {Testi}, {Palla}, \&
  {Zinnecker}}]{deWit2005}
{de Wit}, W.~J., {Testi}, L., {Palla}, F., \& {Zinnecker}, H. 2005, \aap, 437,
  247

\bibitem[{{Evans} \& {Howarth}(2008)}]{EvansHowarth2008}
{Evans}, C.~J., \& {Howarth}, I.~D. 2008, \mnras, 386, 826

\bibitem[{{Gies}(1987)}]{Gies1987}
{Gies}, D.~R. 1987, \apjs, 64, 545

\bibitem[{{Guthrie}(1984)}]{Guthrie1984}
{Guthrie}, B.~N.~G. 1984, \mnras, 210, 159

\bibitem[{{Henize}(1956)}]{Henize1956}
{Henize}, K.~G. 1956, \apjs, 2, 315

\bibitem[{{Hoogerwerf} et~al.(2001){Hoogerwerf}, {de Bruijne}, \& {de
  Zeeuw}}]{Hoogerwerf2001}
{Hoogerwerf}, R., {de Bruijne}, J.~H.~J., \& {de Zeeuw}, P.~T. 2001, \aap, 365,
  49

\bibitem[{{Huang} \& {Gies}(2008)}]{HuangGies2008}
{Huang}, W., \& {Gies}, D.~R. 2008, \apj, 683, 1045

\bibitem[{{Huang} et~al.(2010){Huang}, {Gies}, \& {McSwain}}]{Huang2010}
{Huang}, W., {Gies}, D.~R., \& {McSwain}, M.~V. 2010, \apj, 722, 605

\bibitem[{{Keller} et~al.(1999){Keller}, {Wood}, \& {Bessell}}]{Keller1999}
{Keller}, S.~C., {Wood}, P.~R., \& {Bessell}, M.~S. 1999, \aaps, 134, 489

\bibitem[{{Kroupa}(2001)}]{Kroupa2001}
{Kroupa}, P. 2001, \mnras, 322, 231

\bibitem[{{Krumholz} \& {McKee}(2008)}]{KrumholzMcKee2008}
{Krumholz}, M.~R., \& {McKee}, C.~F. 2008, \nat, 451, 1082

\bibitem[{{Lamb} et~al.(2010){Lamb}, {Oey}, {Werk}, \& {Ingleby}}]{Lamb2010}
{Lamb}, J.~B., {Oey}, M.~S., {Werk}, J.~K., \& {Ingleby}, L.~D. 2010, \apj,
  725, 1886

\bibitem[{{Leonard} \& {Duncan}(1988)}]{LeonardDuncan1988}
{Leonard}, P.~J.~T., \& {Duncan}, M.~J. 1988, \aj, 96, 222

\bibitem[{{Leonard} \& {Duncan}(1990)}]{LeonardDuncan1990}
--- 1990, \aj, 99, 608

\bibitem[{{Maeder} et~al.(1999){Maeder}, {Grebel}, \&
  {Mermilliod}}]{Maeder1999}
{Maeder}, A., {Grebel}, E.~K., \& {Mermilliod}, J.-C. 1999, \aap, 346, 459

\bibitem[{{Ma{\'{\i}}z-Apell{\'a}niz} et~al.(2004){Ma{\'{\i}}z-Apell{\'a}niz},
  {Walborn}, {Galu{\'e}}, \& {Wei}}]{Maiz2004}
{Ma{\'{\i}}z-Apell{\'a}niz}, J., {Walborn}, N.~R., {Galu{\'e}}, H.~{\'A}., \&
  {Wei}, L.~H. 2004, \apjs, 151, 103

\bibitem[{{Martayan} et~al.(2006){Martayan}, {Fr{\'e}mat}, {Hubert}, {Floquet},
  {Zorec}, \& {Neiner}}]{Martayan2006}
{Martayan}, C., {Fr{\'e}mat}, Y., {Hubert}, A.-M., {Floquet}, M., {Zorec}, J.,
  \& {Neiner}, C. 2006, \aap, 452, 273

\bibitem[{{Martayan} et~al.(2007){Martayan}, {Fr{\'e}mat}, {Hubert}, {Floquet},
  {Zorec}, \& {Neiner}}]{Martayan2007}
--- 2007, \aap, 462, 683

\bibitem[{{Martin}(2006)}]{Martin2006}
{Martin}, J.~C. 2006, \aj, 131, 3047

\bibitem[{{Mason} et~al.(2009){Mason}, {Hartkopf}, {Gies}, {Henry}, \&
  {Helsel}}]{Mason2009}
{Mason}, B.~D., {Hartkopf}, W.~I., {Gies}, D.~R., {Henry}, T.~J., \& {Helsel},
  J.~W. 2009, \aj, 137, 3358

\bibitem[{{Massey}(2002)}]{Massey2002}
{Massey}, P. 2002, \apjs, 141, 81

\bibitem[{{Massey} et~al.(2001){Massey}, {DeGioia-Eastwood}, \&
  {Waterhouse}}]{Massey2001}
{Massey}, P., {DeGioia-Eastwood}, K., \& {Waterhouse}, E. 2001, \aj, 121, 1050

\bibitem[{{Massey} \& {Duffy}(2001)}]{MasseyDuffy2001}
{Massey}, P., \& {Duffy}, A.~S. 2001, \apj, 550, 713

\bibitem[{{Massey} et~al.(1995){Massey}, {Johnson}, \&
  {Degioia-Eastwood}}]{Massey1995}
{Massey}, P., {Johnson}, K.~E., \& {Degioia-Eastwood}, K. 1995, \apj, 454, 151

\bibitem[{{Mikkola}(1983)}]{Mikkola1983}
{Mikkola}, S. 1983, \mnras, 205, 733

\bibitem[{{Moffat}(1998)}]{Moffat1998}
{Moffat}, A.~F.~J. e.~a. 1998, \aap, 331, 949

\bibitem[{{Oey} et~al.(2004){Oey}, {King}, \& {Parker}}]{Oey2004}
{Oey}, M.~S., {King}, N.~L., \& {Parker}, J.~W. 2004, \aj, 127, 1632

\bibitem[{{Poveda} et~al.(1967){Poveda}, {Ruiz}, \& {Allen}}]{Poveda1967}
{Poveda}, A., {Ruiz}, J., \& {Allen}, C. 1967, Bol. Obs. Tonantzintla y
  Tacubaya, 4, 86

\bibitem[{{Sana} \& {Evans}(2011)}]{SanaEvans2011}
{Sana}, H., \& {Evans}, C.~J. 2011, in IAU Symposium 272, edited by {C.~Neiner,
  G.~Wade, G.~Meynet, \& G.~Peters}, 474

\bibitem[{{Slettebak} et~al.(1997){Slettebak}, {Wagner}, \&
  {Bertram}}]{Slettebak1997}
{Slettebak}, A., {Wagner}, R.~M., \& {Bertram}, R. 1997, \pasp, 109, 1

\bibitem[{{Strom} et~al.(2005){Strom}, {Wolff}, \& {Dror}}]{Strom2005}
{Strom}, S.~E., {Wolff}, S.~C., \& {Dror}, D.~H.~A. 2005, \aj, 129, 809

\bibitem[{{van den Bergh}(2004)}]{vandenBergh2004}
{van den Bergh}, S. 2004, \aj, 128, 1880

\bibitem[{{Wisniewski} et~al.(2007){Wisniewski}, {Bjorkman}, {Bjorkman}, \&
  {Clampin}}]{Wisniewski2007}
{Wisniewski}, J.~P., {Bjorkman}, K.~S., {Bjorkman}, J.~E., \& {Clampin}, M.
  2007, \apj, 670, 1331

\bibitem[{{Wolff} et~al.(2008){Wolff}, {Strom}, {Cunha}, {Daflon}, {Olsen}, \&
  {Dror}}]{Wolff2008}
{Wolff}, S.~C., {Strom}, S.~E., {Cunha}, K., {Daflon}, S., {Olsen}, K., \&
  {Dror}, D. 2008, \aj, 136, 1049

\bibitem[{{Wolff} et~al.(2007){Wolff}, {Strom}, {Dror}, \& {Venn}}]{Wolff2007}
{Wolff}, S.~C., {Strom}, S.~E., {Dror}, D., \& {Venn}, K. 2007, \aj, 133, 1092

\bibitem[{{Zwicky}(1957)}]{Zwicky1957}
{Zwicky}, F. 1957, {Morphological Astronomy} (Berlin: Springer, 1957)

\end{thebibliography}

\end{document}